**The high gas pressure HIP influence on structure and transport properties of MgB2 superconductors of single and multicore composition.**

T. Cetner[1], A. Morawski[1], M. Rindfleisch[2], M. Tomsic[2], A. Presz[1], D. Gajda[3], A. Zaleski[4], O. Tkachenko[4]

1 Institute of High Pressure Physics, Polish Academy of Sciences, Sokołowska 29/37, 01-142 Warszawa. Poland
2 Hyper Tech Research, Inc., 1275 Kinnear Road, Columbus, OH 43212 United States,
3 International Laboratory of High Magnetic Fields and Low Temperature Gajowicka 95, 53-421 Wrocław, Poland
4 Instytut of Low Temperature and Structure Research Polish Academy of Sciences, Okólna 2, 50-422 Wrocław, Poland

Superconducting $MgB_2$ wires in Cu, GlidCop or Monel sheath with Nb or Fe barrier are prepared. Wires vary by sheath material, number of superconducting cores and their chemical composition. Wires are HIP-ed (Hot Isostatic Pressing) at various temperatures ($600-800$) and pressure (up to $1.4 GPa$). SEM pictures of cross sections are investigated in order to investigate barrier reactivity and cracking, superconducting material density and grain sizes. Transport measurements are made in magnetic field up to $14T$ leading to calculations of critical current density $j_c$ and global pinning force $F_p$. Improvement of transport properties due to higher density of superconducting material is shown.

INTRODUCTION

Since discovery of its superconducting properties in 2001 [1], MgB2 became one of the most promising materials for preparation of superconducting wires [2]. Apart from its high critical properties, MgB2 has such advantages as a simple chemical composition of two ingredients, physical form of powder that allows easy shaping of the material and low prices for boron and magnesium.

Preparation of wires

Wires presented here were prepared in the PIT (Powder In Tube) technique by [3] using an unique and patented method of continuous tube forming and filling (CTFF) [4]. Key feature of this method involves shaping of a flat metal strip with dies, filling it with powder while it is in U-shape to finally close it and obtain a wire (see Figure 1).

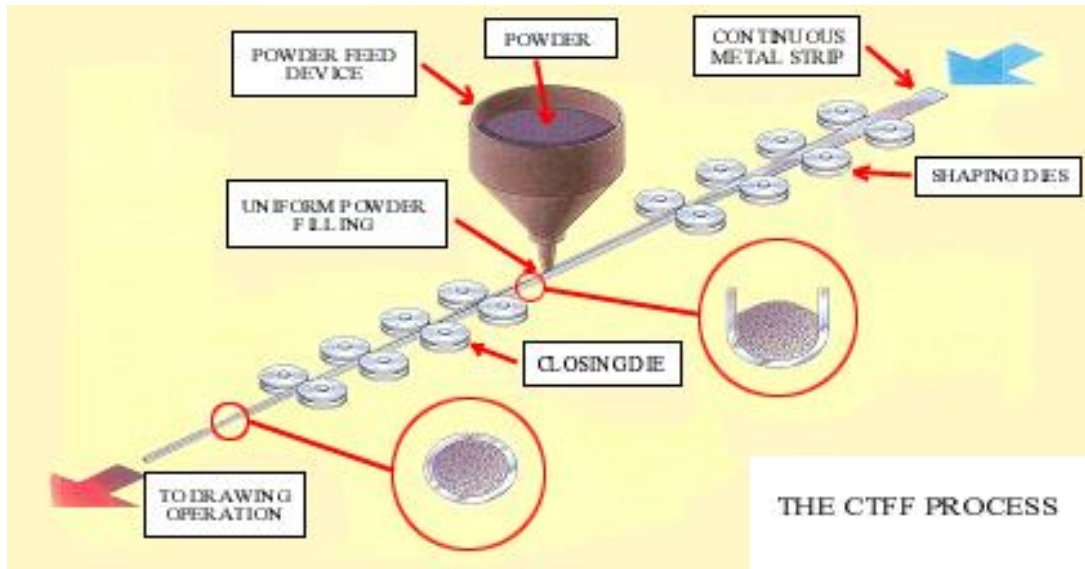

Figure 1 Scheme of CTFF process: continuous forming and filling of superconducting wires

Important issue in production of $MgB_2$ wires is the chemical composition of the initial powder [5]. Apart from high purity boron and magnesium, various additives are used to increase critical properties, especially in high magnetic fields. Powder compositions used here are presented in Table 1.

Properties of the sheath material, its thermal and electrical conductivity as well as plasticity and hardness, influence mechanical properties of the whole wire, like its capability of being drawn and its behaviour when in non-superconducting state. As the desired properties of the sheath material are low electrical resistivity and high thermal conductivity, copper and Monel (a Cu and Ni alloy) are preferred over iron. However, due to high reactivity of copper with magnesium, one has to introduce a chemical barrier between inner superconducting core and the sheath. Here two kinds of barrier were used: niobium and iron. Multi-core wires are prepared by enclosing each core in the barrier material, than in the sheath material and finally placing all cores in a common outer sheath. Such wires are than drawn to desired diameters (see Table 1).

Table 1  Structure and chemical composition of wires annealed through HIP

| wire ID | no. of cores | barrier | core sheath | outer sheath | B source | Mg to B ratio | additive | d [mm] | fill factor [%] |
|---|---|---|---|---|---|---|---|---|---|
| 03 | 6 | Nb | Cu | Monel | SMI | 1:2 | C | 0.83 | 17.7 |
| 18 | 6 | Nb | Cu | Cu | 99B | 1.10:2 | SiC | 0.83 | 14.9 |
| 22 | 18 | Fe | Cu | Glidcop | Ts | 1:2 | C4H6O3 | 0.83 | 13.9 |
| 30 | 6 | Nb | Cu | Monel | 99B | 1.10:2 | SiC | 0.83 | 15.0 |
| 43 | 18 | Nb | Cu | Monel | 99B | 1.10:2 | - | 0.83 | 15.0 |
| 70 | 1 | Fe | Cu | - | Ts | 1:2 | C4H6O3 | 0.83 | 28.7 |
| 76 | 6 | Nb | Cu | Glidcop | 99B | 1.10:2 | SiC | 0.83 | 16.6 |
| 92 | 6 | Nb | Cu | Cu | 99B | 1.10:2 | - | 0.83 | 19.4 |

Thus prepared wires are typically heated to ca. $700^oC$ in order to cause reaction between magnesium and boron and form the superconducting material. During this process it is important to obtain high density and small grain sizes in the $MgB_2$ phase. It is a major technological challenge, as during this reaction system shrinks by up to 30% of initial volume size. In this work, we use HIP (Hot Isostatic Pressing) to press wires during reaction and provide high density of superconducting material. HIP was applied in high purity argon atmosphere in a pressure chamber (see Figure 2) with highest gas pressure of $1.4 GPa$.

Dynamics of reaction process is driven by the gas pressure, temperature and time of annealing, thus one has to tune all those parameters to obtain optimal grain structure. Various settings of HIP used here are presented in Table 2.

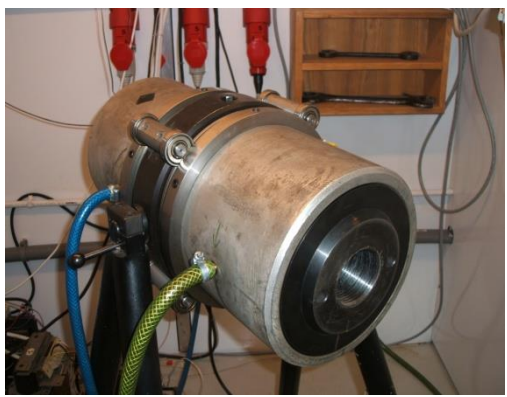

Figure 2 Pressure chamber used for HIP

Table 2  Parameters of HIP applied to every wire sample

| process ID | temperature [°C] | pressure [kbar] | time [h:min] |
|---|---|---|---|
| I | 700 | 0,001 | 0:15 |
| II | 700 | 0,01 | 0:15 |
| III | 700 | 0,2 | 1:00 |
| IV | 700 | 4 | 0:30 |
| V | 600 | 8 | 12:00 |
| VI | 700 | 10 | 0:15 |
| VII | 700 | 14 | 0:30 |

SEM RESULTS

Powder density

A significant effect of pressing the wires during annealing is visible on SEM pictures (see Figure 3). For samples that reacted in lower pressure one can notice numerous gaps in superconducting material, caused by shrinking of the material during reaction. However, for sample after HIP in $1.4 GPa$ such gaps are very rare, thus causing higher density of superconducting material.

Grain sizes

As already mentioned, grain sizes are an important characteristic of the superconducting material, affecting critical properties of wires. Here investigation of the grain sizes is performed through SEM pictures (see Figure 4). There is no significant influence of HIP pressure on grain sizes, which are typically in range of $50\text{nm} \div 100\text{nm}$. However, there is a visible improvement for a long process ($12h$) in a relatively low temperature ($600^oC$), where grains are of a much smaller size. Unfortunately, there is no information on the grain connectivity with this method.

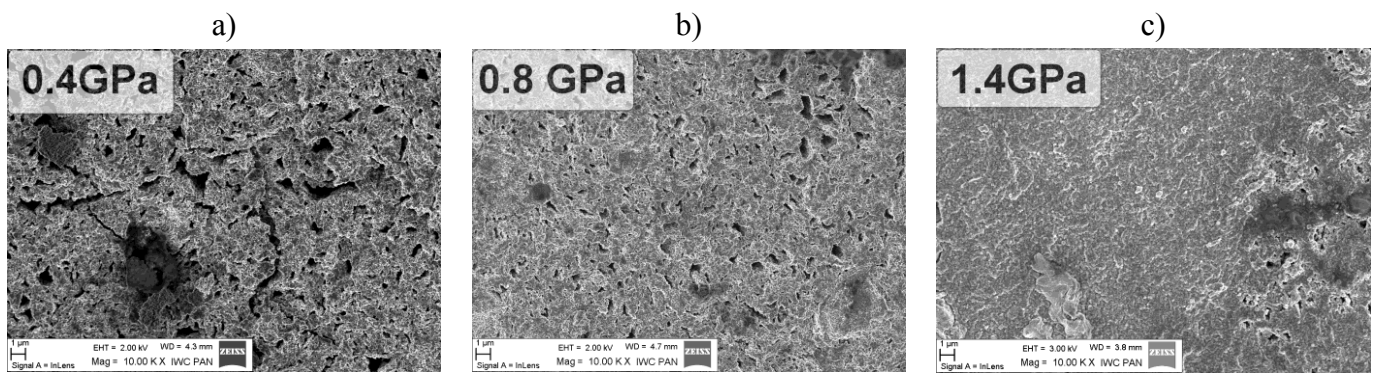

Figure 3  SEM pictures of wire 43 after HIP in various gas pressures (increasing to the right side), that show differences in powder density. Sample *a* is from process no. IV, *b* from V and *c* from VII.

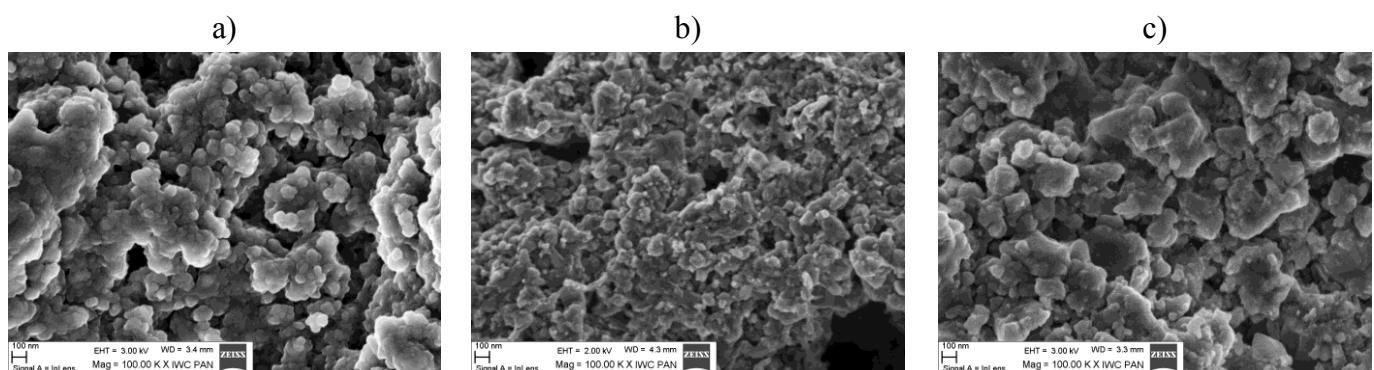

Figure 4  SEM pictures of wire 22 after HIP in various conditions. Sample *a* is from process no. I, *b* from V and *c* from VII.

Barrier quality

Preparation of wires, including barrier parameters, was optimised for annealing under atmospheric pressure, where barrier worked properly. However, it is visible on SEM pictures that for HIP one has to provide a higher quality of the barrier. Minor defects in barrier under low pressure effect in a local reaction between core and sheath, leaving majority of superconducting material unharmed. On the contrary, under high pressure sheath material is pressed into core even through smallest cracks in the barrier (see Figure 5), what results in total destruction of the given superconducting core.

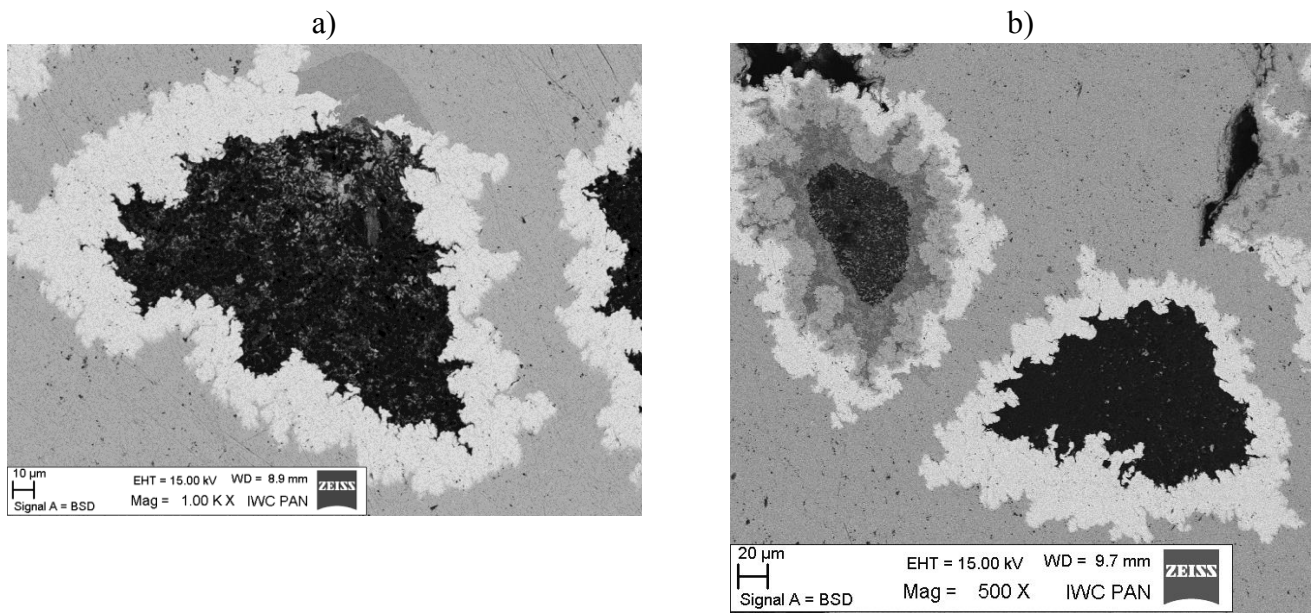

Figure 5  SEM pictures of wire 18 after HIP no. I (a) and VII (b). On the left picture there is a break in the barrier that, being in low pressure, caused only a local reaction with the sheath , on the right side one of the cores has fully reacted with the sheath material as an effect of barrier defect and high pressure.

CRITICAL CURRENT MEASUREMENTS

Several transport measurements were done in Wrocaław[3] in a magnetic field up to 14T, produced by a Bitter magnet, with a maximum current of 150A [6]. Some cross-check measurements were additionally done in Wien [7], where it was possible to apply higher current to the samples. In general, measurements showed that annealing through HIP can significantly improve critical properties of wires. It is however sensible to the composition of wires and for part of the samples there was no improvement, or even a decrease of critical current. Chosen results of those measurements are presented.

Improvement of critical properties for wire 43 was confirmed by independent measurements (see Figure 6), although $j_C$ results are different. This could be caused by defects in barrier in samples measured in Wien. Calculations of pinning force show a significant increase. Maximal $F_P$ is at very low B for HIP-ed sample, which suggests inter-grain mechanism of pinning [8].

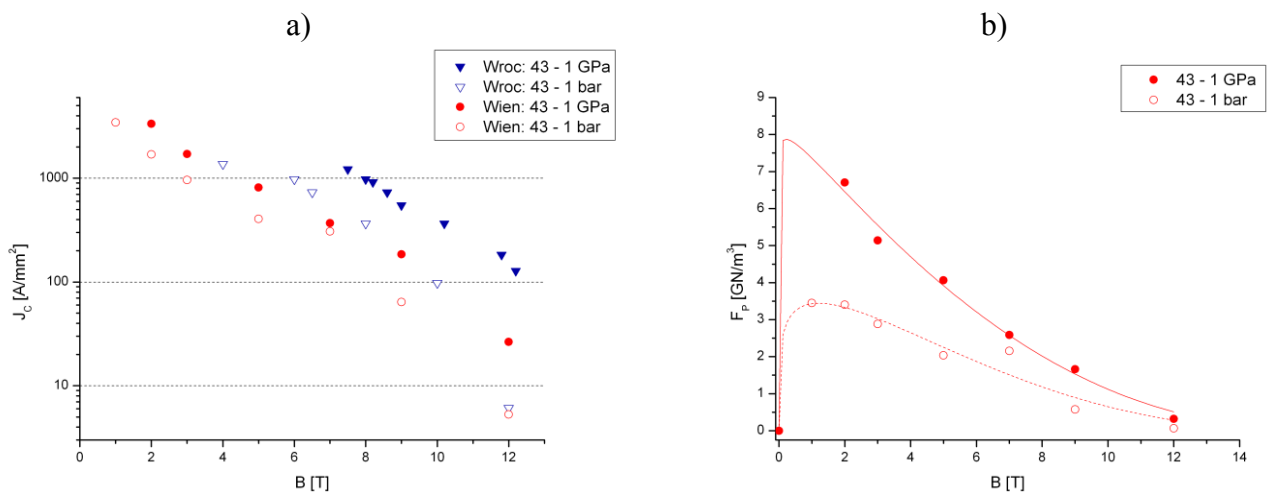

Figure 6   Results of current measurements for wire 43 performed at two different laboratories. Increase of $j_C$ (a) is confirmed by both of them. $F_P$ could be calculated in a wide B range only for one measurement and showed a significant increase (b)

Interesting results are obtained for wires with SiC addition. For both available samples an improvement was observed at high magnetic field, together with a decrease at lower field (see Figure 7). This is clearly visible in the pinning force dependence on B. For both samples, the maximum $F_P$ is obtained at higher magnetic field for sample annealed in gas pressure of 1GPa. Critical current for wire 30 annealed at 1bar pressure was not measured at low B, due to limitation of maximal current in a wire.

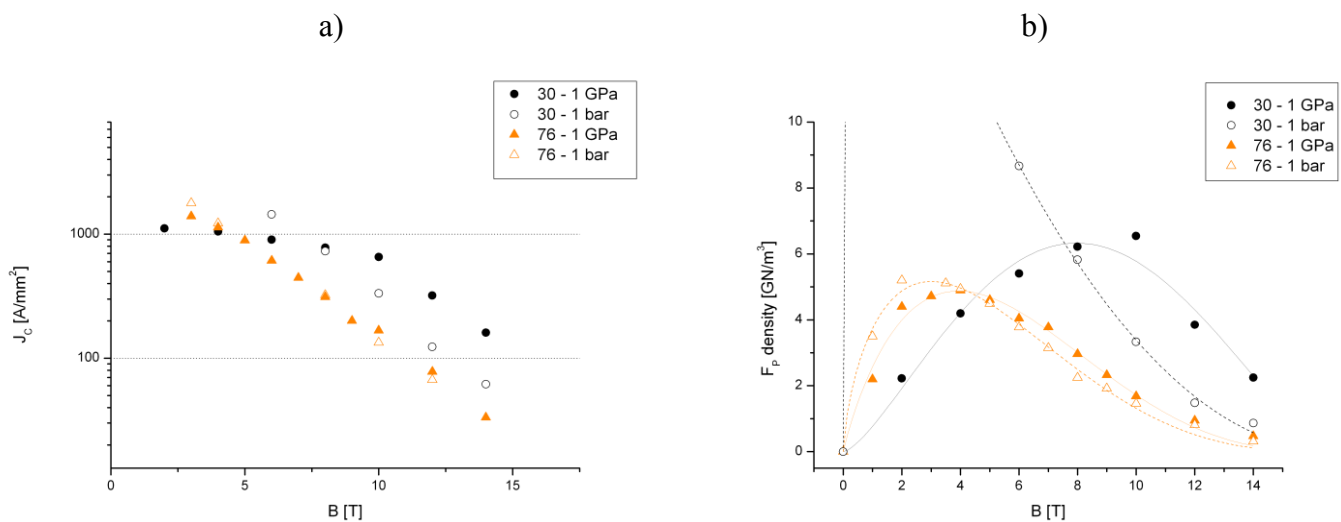

Figure 7   Results of current measurements for two different samples with SiC addition after annealing in different gas pressure. Plot of $j_C$ vs. B (a) and of $F_p$ density vs. B (b) show that SiC addition causes shift in the pinning force towards higher magnetic fields, thus improving wire behaviour in this region

CONCLUSIONS

Superconducting $MgB_2$ wires of various sheath/barrier material and additives were annealed at various conditions, mostly to evaluate influence of high gas pressure up to 1.4GPa available during HIP.

High pressure proved to increase density of superconducting core that can lead to improvement in critical properties. Significance of barrier quality at higher pressure was shown with SEM pictures, where a minor defect in barrier caused reaction in a whole superconducting core. HIP improved critical current density and pinning force for many of used wires. Addition of SiC improved wire parameters in high magnetic field.

ACKNOWLEDGEMENTS

This work was partially financed by Polish Ministry of Science and Information, Technology Project no. N N510 278738 and Project no. N N510 35747.